\documentclass[twocolumn,a4paper,accepted=2019-06-13]{quantumarticle}
\usepackage[numbers,sort&compress]{natbib}

\usepackage[utf8]{inputenc}
\usepackage[english]{babel}
\usepackage[T1]{fontenc}

\usepackage{amsmath}
\usepackage{amsfonts}
\usepackage{amssymb}
\usepackage{amsthm}
\usepackage{siunitx}
\usepackage{dsfont}
\usepackage{microtype}

\newcommand{\cf}{cf.\ }

\newcommand{\coloneq}{\mathrel{\mathop:}=}
\newcommand{\eqcolon}{=\mathrel{\mathop:}}
\newcommand{\dd}{\mathrm{d}}
\newcommand{\ew}[1]{\left\langle{#1}\right\rangle}

\newcommand{\kB}{k_\mathrm{B}}
\newcommand{\sminus}{\sigma_-}
\newcommand{\splus}{\sigma_+}
\newcommand{\indexc}{\mathrm{c}}
\newcommand{\indexh}{\mathrm{h}}
\newcommand{\indexWM}{\mathrm{wm}}
\newcommand{\omegac}{\mathcal{E}_\indexc}
\newcommand{\omegah}{\mathcal{E}_\indexh}
\newcommand{\omegaT}{\omega_\mathrm{T}}
\newcommand{\Tc}{T_\indexc}
\newcommand{\Th}{T_\indexh}
\newcommand{\nbar}{\bar{n}}
\newcommand{\mWM}{m_\indexWM}
\newcommand{\NWM}{N_\indexWM}
\newcommand{\lambdaWM}{\lambda_\indexWM}
\newcommand{\gammaWM}{\gamma_\indexWM}
\newcommand{\tauStroke}{\tau_\mathrm{stroke}}

\begin{document}

\title{Quantized refrigerator for an atomic cloud}

\author{Wolfgang Niedenzu}
\affiliation{Institut f\"ur Theoretische Physik, Universit\"at Innsbruck, Technikerstra{\ss}e~21a, A-6020~Innsbruck, Austria}
\email{Wolfgang.Niedenzu@uibk.ac.at}
\orcid{0000-0001-7122-3330}

\author{Igor Mazets}
\affiliation{Vienna Center for Quantum Science and Technology (VCQ), Atominstitut, TU Wien, 1020 Vienna, Austria}
\affiliation{Wolfgang Pauli Institute, c/o Fakult\"at für Mathematik, Universit\"at Wien, 1090 Vienna, Austria}
\orcid{0000-0002-3769-8313}

\author{Gershon Kurizki}
\affiliation{Department of Chemical Physics, Weizmann Institute of Science, Rehovot~7610001, Israel}

\author{Fred Jendrzejewski}
\affiliation{Heidelberg University, Kirchhoff-Institut f\"ur Physik, Im Neuenheimer Feld 227, D-69120 Heidelberg, Germany}
\email{fnj@kip.uni-heidelberg.de}
\orcid{0000-0003-1488-7901}

\date{June 24, 2019}

\begin{abstract}
  We propose to implement a quantized thermal machine based on a mixture of two atomic species. One atomic species implements the working medium and the other implements two (cold and hot) baths. We show that such a setup can be employed for the refrigeration of a large bosonic cloud starting above and ending below the condensation threshold. We analyze its operation in a regime conforming to the quantized Otto cycle and discuss the prospects for continuous-cycle operation, addressing the experimental as well as theoretical limitations. Beyond its applicative significance, this setup has a potential for the study of fundamental questions of quantum thermodynamics.
\end{abstract}

\maketitle

\section{Introduction}

Over the past two decades, extensive studies of thermal machines in the quantum domain~\cite{alicki1979quantum,kosloff1984quantum,gelbwaser2015thermodynamics,goold2016role,vinjanampathy2016quantum,kosloff2017quantum,ghosh2019thermodynamic} have sought to reveal either fundamentally new aspects of thermodynamics or unique quantum advantages compared to their classical counterparts. Yet, despite the great progress achieved, both theoretically~\cite{scully2003extracting,abah2012single,horodecki2013fundamental,skrzypczyk2014work,brask2015autonomous,uzdin2015equivalence,campisi2016power,niedenzu2018quantum} and experimentally~\cite{brantut2013thermoelectric,koski2014experimental,rossnagel2016single,klaers2017squeezed,klatzow2019experimental,koski2015chip,vonlindenfels2018spin}, the potential of quantum thermal machines for useful quantum technology applications only begins to unfold. In particular, new insights into thermodynamics in the quantum domain may be obtained by extension of these studies to hitherto unexplored quantum phenomena. In this spirit, we suggest to realise thermal machines in cold atomic gases, aiming at their improved refrigeration.

\begin{figure}
  \centering
  \includegraphics[width=\columnwidth]{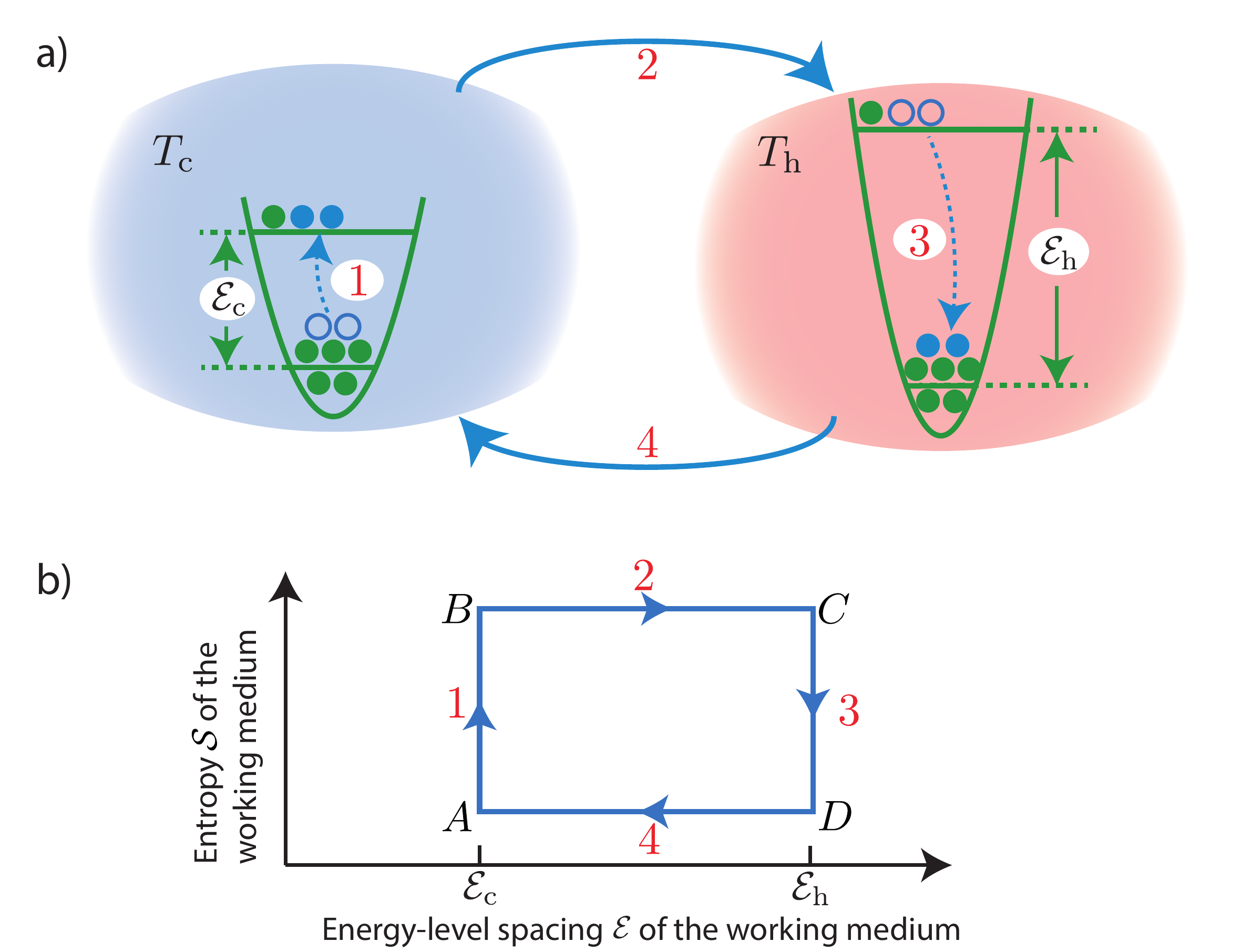}
  \caption{\textbf{Quantized Otto cycle for refrigeration of the cold bath:} a)~Schematic representation of the Otto cooling cycle. The working medium (green) is located in a strongly confining harmonic potential. During the heat-exchange strokes 1 and 3 it is coupled, sequentially to the cold bath (left, blue) and the hot bath (right, red). The energy level spacing of the working medium is adiabatically changed during the strokes 2 and 4. b)~The quantized Otto cycle in the energy--entropy plane of the working medium. The cycle is transversed in the clockwise direction (refrigeration mode) between the end points $A\rightarrow B\rightarrow C\rightarrow D\rightarrow A$.}\label{fig_cycle}
\end{figure}

Specifically, we propose to implement a quantized thermal machine, which is based on a mixture of two cold atomic gases, and employ it as a refrigerator for one of these species. The goal would be to cool the atomic gas starting from the thermal regime and ending well within the quantum-degenerate regime. The proposed cycle will be a fundamentally new avenue for cooling, only limited by thermalization and not by coolant recoil or by atom loss, as do existing schemes, such as laser cooling or evaporative cooling~\cite{pethickbook}. In a simple implementation of the Otto cycle~\cite{geva1992quantum,feldmann2003quantum,abah2016optimal,kosloff2017quantum,erdman2018maximum} (see Fig.~\ref{fig_cycle}) the working medium (WM) is realised by one species that is alternately coupled to spatially separated, hot and cold baths of another species. This allows for progressive cooling of the cold bath at the expense of heat dumping into the hot bath. Traversing the quantized Otto cycle several hundred times should then allow us to reduce the temperature of the cold bath by an order of magnitude and even cross the phase-transition threshold towards Bose--Einstein condensation. The proposed cooling technique should also be applicable to other types of baths, including strongly correlated Mott insulators or anti-ferromagnetic phases, where cooling by existing techniques remains an experimental challenge~\cite{mazurenko2016cold}. We note that in our refrigeration setup there is a clear physical distinction between the WM and the two thermal baths (cf.\ Fig.~\ref{fig_cycle}), contrary to the engine proposed in Ref.~\cite{fialko2012isolated}.

\par

A multitude of diverse cooling techniques have revolutionized atomic physics over the last thirty years. Conventional laser cooling that has founded the field of ultracold atomic gases employs photon recoil for heat extraction from the atomic cloud. However, the phase-space-density is fundamentally limited by the recoil energy of the photon~\cite{pethickbook}. Degenerate quantum gases have been realized by evaporative cooling, whereby the highest-energy atoms are selectively removed from the cloud~\cite{davis1995evaporative,petrich1995stable}. These evaporative techniques are fundamentally limited by the required spatial separation of regions possessing high or low entropy. Such separation typically breaks down in the quantum degenerate regime. None of these limitations should apply to the Otto cycle, which we will study now.

\par

\section{Otto cycle for cold quantum gases}

The four strokes of the quantized Otto cycle can be implemented as follows in cold-atom setups. The working medium (WM) is formed by a tightly trapped atomic gas, such that only two quantum states of the confined system are accessible. Such a configuration arises naturally in highly focused optical traps~\cite{grimm2000optical}. It can be modelled by the Hamiltonian $H=\mathcal{E}\splus\sminus$, where $\sminus$ ($\splus$) are the Pauli lowering (raising) operators in the effective two-level trap. The spacing between the energy levels is varied by an external optical laser which provides the work input required for refrigeration, analogous to a piston in common thermodynamic machine cycles~\cite{alicki1979quantum,kosloff1984quantum,geva1992quantum,kolar2012quantum,gelbwaser2013minimal}.

\par

In the first stroke, the WM has energy-level spacing $\omegac$ and thermalizes through its interaction with a thermal bath at temperature $\Tc$. Thermalization is ensured through contact collisions between the atoms of the bath and the WM. Such thermalization does not involve any radiative processes, which limit the maximal rate of laser cooling. The proposed thermalization represents the isochoric interaction with the cold bath of the Otto cycle (stroke~1 in Fig.~\ref{fig_cycle}). The cold bath is realized by an atomic species that differs from the WM (Fig.~\ref{fig_cycle}a), modelled here as an ideal, uncondensed Bose gas. However, the same cycle remains valid for more intricate baths, such as Mott insulators, superfluids or fermionic (spin) baths. During this first stroke the WM receives from the cold bath the heat $Q_\indexc=\omegac(\nbar_\indexc-\nbar_\indexh)$, where $\nbar_i$ ($i\in\{\indexc,\indexh\}$) are the thermal excitations of the two-level WM (see Appendix~\ref{app_otto}).

\par

In the second stroke, the WM is adiabatically decoupled from the cold bath and its energy-level spacing is adiabatically raised to $\omegah>\omegac$. This realizes the isentropic compression stroke of the Otto cycle (stroke~2 in Fig.~\ref{fig_cycle}). The work $W_\mathrm{in}=(\omegah-\omegac)\nbar_\indexh$ which is needed for the WM compression is extracted from the classical optical field which confines the WM. We here neglect the work necessary for the compression of the atoms that remain in the ground state as it cancels with the decompression in the last stroke of the cycle.

\par

In the third stroke, the WM is coupled to the hot thermal bath at temperature $\Th>\Tc$, with which the WM thermalizes (Stroke~3 in Fig.~\ref{fig_cycle}). This stroke realizes the second isochore in the Otto cycle in which the WM deposits the heat $Q_\indexh=\omegah(\nbar_\indexh-\nbar_\indexc)$ into the hot bath.

\par

The fourth stroke that closes the cycle consists in adiabatic decoupling of the WM from the hot bath and adiabatic decrease of the energy-level spacing back to $\omegac$ (stroke~4 in Fig.~\ref{fig_cycle}). This decompression of the WM produces the work $W_\mathrm{out}=(\omegac-\omegah)\nbar_\indexc$ which is transferred to the optical field.

\par

The Otto cycle acts as a refrigerator for the cold bath as long as $Q_\indexc>0$ ($Q_\indexh<0$), which yields the cooling condition
\begin{equation}\label{eq_otto_cooling_condition}
  \nbar_\indexc>\nbar_\indexh\quad\Rightarrow\quad\frac{\omegah}{\Th}>\frac{\omegac}{\Tc}.
\end{equation}
The theoretically-achievable minimal temperature of the cold bath is then limited by condition~\eqref{eq_otto_cooling_condition}, yielding
\begin{equation}\label{eq_otto_Tc_min}
  \Tc^\mathrm{min}=\frac{\omegac}{\omegah}\Th.
\end{equation}
The ratio of the two baths temperature is hence only limited by the tunability of the energy-level spacing of the quantized WM. For an optically confined WM, the energy-level spacing $\mathcal{E}$ is controlled by the square root of the confining laser intensity $I$, $\mathcal{E} \propto \sqrt{I}$~\cite{grimm2000optical}. Since intensity ratios of $I_\indexh/I_\indexc\gtrsim 100$ are experimentally controllable, it should be feasible to refrigerate the cold-bath temperature by an order of magnitude.

\par

\section{Proposed experimental implementation}

\begin{figure}
  \centering
  \includegraphics[width=\columnwidth]{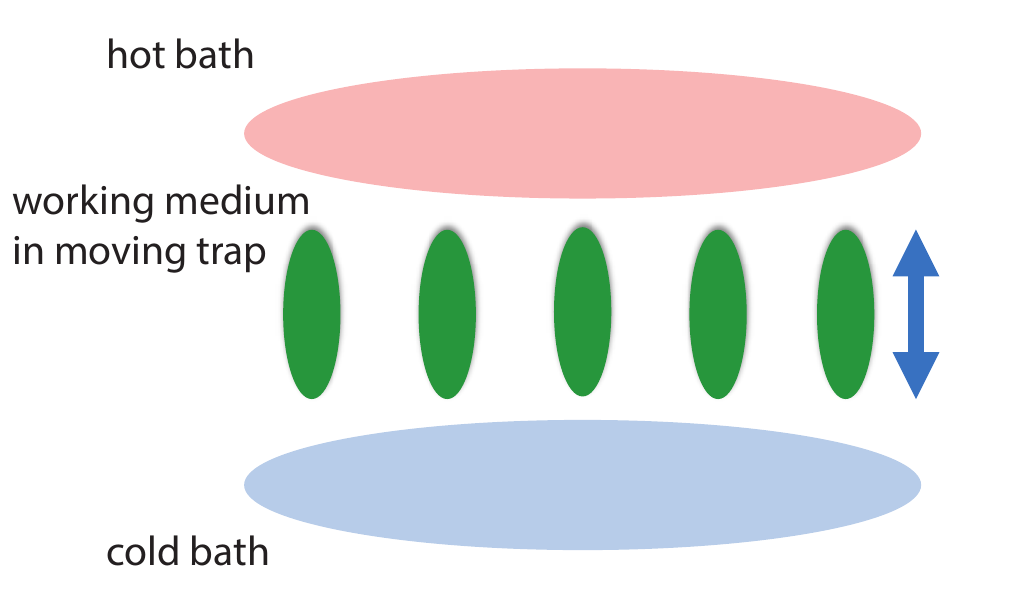}
  \caption{\textbf{Envisaged implementation of the Otto cycle:} The two baths are formed by large clouds of an atomic species. The working medium is implemented by another strongly confined atomic species. Alternate coupling of the working medium to each bath (consecutively) is achieved by spatially moving the working medium into the bath region.}\label{fig_experiment}
\end{figure}

The quantized Otto refrigerator can be implemented by employing the following state-of-the-art techniques: (i)~The spatially separated hot and cold baths can be created within two separate harmonic traps, as originally employed in atomic tunneling junctions~\cite{albiez2005direct} or more recently in atomic circuits with fermionic and bosonic baths~\cite{brantut2013thermoelectric,eckel2016contact}. In the proposed setup (see Fig.~\ref{fig_experiment}) the two baths must be completely separated such that no particle exchange or heat flow can happen between them, except via the WM.

\par

\par

(ii)~The WM is formed by an atomic species, by means of a species-selective optical lattice or tweezer. Such schemes have been employed for single bosonic atoms immersed in a cold gas~\cite{spethmann2012dynamics,hohmann2016single} or for larger numbers of fermionic and bosonic atoms immersed in a Bose--Einstein condensate (BEC)~\cite{scelle2013motional,rentrop2016observation}.

\par

(iii)~Control of energy-level spacing and position of the WM can be implemented by standard optical techniques~\cite{bloch2005ultracold,schymik2018implementing}.

\par

(iv)~The temperature of each bath has to be measured independently through \textit{in-situ} measurements of the bath profile and density correlations~\cite{ku2012revealing,jacqmin2011subpoissonian,desbuquois2014determination}. Another intriguing approach would be to employ an impurity as a local thermometer~\cite{hohmann2016single,correa2017enhancement,lampo2017bose,mukherjee2017enhanced}, provided the thermalization between the bath and the WM is well at hand~\cite{gring2012relaxation}.

\par

\section{Modeling the performance}

In order to quantitatively evaluate the performance of the described cooling technique, we have simulated the Otto cycle for realistic experimental parameters~\cite{scelle2013motional,hohmann2016single}. The simulation considers here bosonic baths formed by Cs atoms and a WM formed by Rb atoms. Initially, both baths are thermal clouds at temperature $\Tc=\Th=\SI{1}{\micro \kelvin}$. Both Cs clouds are confined in a weak trap of frequency $\omega_\mathrm{T}=2\pi\times \SI{80}{\hertz}$ for the cold bath and $\omega_\mathrm{T}=2\pi\times \SI{150}{\hertz}$ for the hot bath. The baths consist of $N_\mathrm{at}^\indexc=2\times 10^5$ and $N_\mathrm{at}^\indexh=50\times 10^5$ atoms, respectively, and these numbers are conserved throughout refrigeration. The initial temperatures of both baths are well above the BEC critical temperature, namely, the bath are initially thermal clouds. The critical temperatures for condensation of the two baths are $T_\mathrm{crit}^\indexc\approx 395\,\mathrm{nK}$ and $T_\mathrm{crit}^\indexh\approx 617\,\mathrm{nK}$, respectively, well within the feasible refrigeration range.

\par

After each cycle, the two bath temperatures are adapted as follows ($i\in\{\indexc,\indexh\}$),
\begin{equation}\label{eq_temperature_change}
  T_i(n+1)=T_i(n)-\frac{\NWM Q_i(n)}{C_\mathrm{V}(T_i(n),N_\mathrm{at}^i)},
\end{equation}
where $\NWM$ is the number of atoms in the WM. The explicit form of the heat capacity $C_\mathrm{V}$ is given in Appendix~\ref{App:BecProperties}. The heat removed from the cold bath in the $n$th cycle is $-\NWM Q_c(n)$, where the negative sign indicates the direction of the heat flow from the cold bath to the WM.

\par
\begin{figure}
  \centering
  \includegraphics[width=\columnwidth]{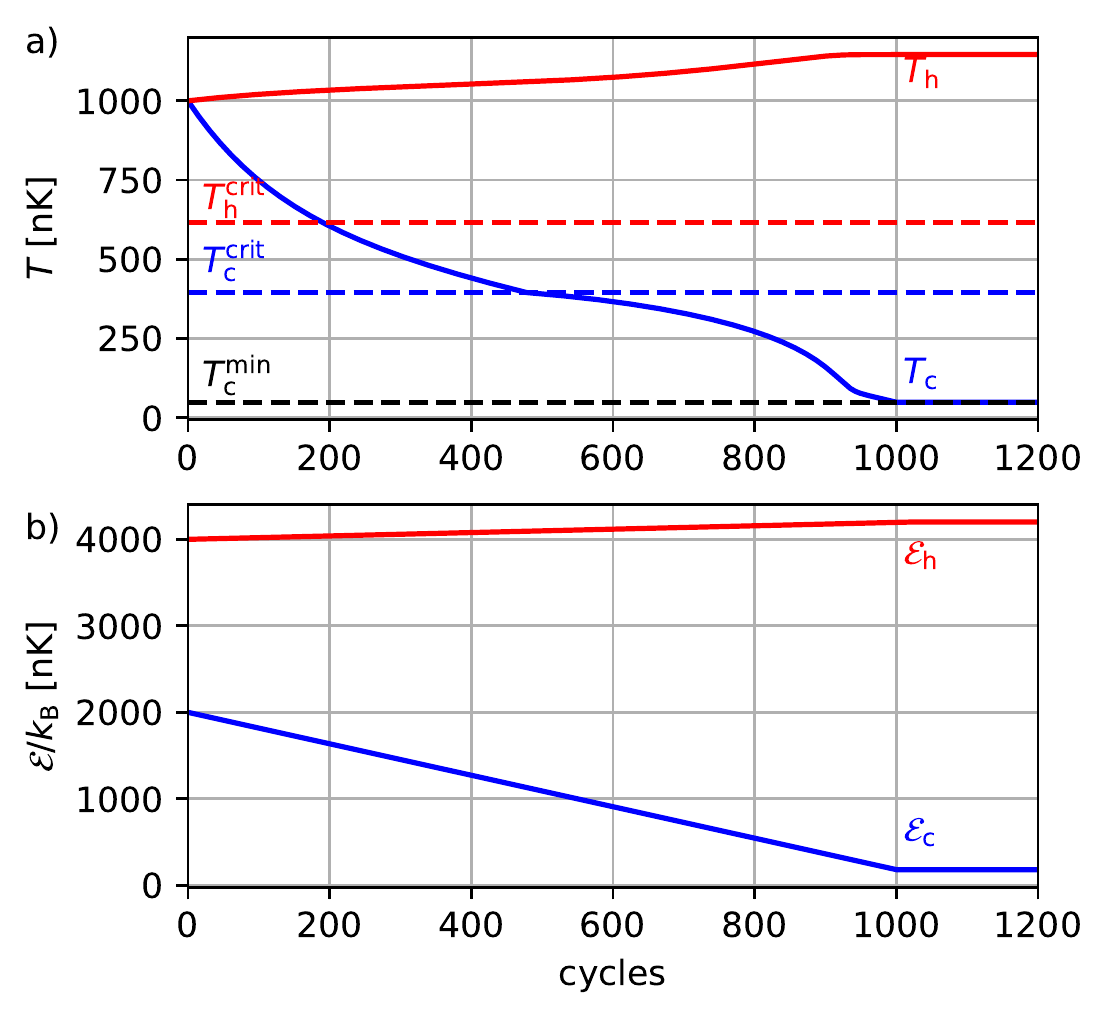}
  \caption{\textbf{Temperatures of the two atomic clouds:} Otto cooling of an initially thermal atomic cloud that serves as a cold bath at temperature $\Tc$ in the proposed refrigerator. The up- and down-ramps on the energy-level spacing have here been terminated after 1000 cycles, consistently with limitations on the total cooling time and tunability of the energy-level spacing. See text for parameters and details.}\label{fig_temperatures_above_Tcrit_h_Tc_700}
\end{figure}
\par

The energy-level spacing of the WM, taken to consist of $\NWM=10^4$ particles, is initially set during the cold and hot isochores to $\omegac/\kB=\SI{2}{\micro\kelvin}$ and $\omegah/\kB=\SI{4}{\micro\kelvin}$, respectively. This choice ensures that only the first excited level of the WM has a non-negligible occupancy, so that heat transport between the two baths can occur according to Eqs.~\eqref{eq_otto_cooling_condition} and~\eqref{eq_otto_Tc_min}. During the Otto cycle we have ramped the energy-level spacings up and down in a linear fashion (see Fig.~\ref{fig_temperatures_above_Tcrit_h_Tc_700}b) such that the occupancy of the upper WM level is always a few percent. The up- and down-ramps have been terminated after 1000 cycles, when the temperature of the cold bath becomes comparable to the mode spacing in the cold bath, $\kB \Tc^\mathrm{min} \approx 7 \hbar \omega_\mathrm{T}$. The reason for this choice is our wish to avoid non-Markovian effects, which may become prominent as the number of occupied bath modes becomes small (see below). We note that the minimal temperature is calculated according to Eq.~\eqref{eq_otto_Tc_min} from the final energy-level spacings and the final temperature of the hot bath.

\par

It can be seen in Fig.~\ref{fig_temperatures_above_Tcrit_h_Tc_700} that efficient cooling takes place for as many cycles as needed to achieve the minimal allowed temperature. After approximately 500 cycles, the condensation threshold of the cold bath is crossed and efficient cooling continues well into the deeply degenerate regime, down to $T\approx 0.1 T_\mathrm{crit}^\indexc$, a regime that is hard to reach by other methods~\cite{olf2015thermometry}. Assuming a cycle time of roughly $\SI{10}{\milli \second}$ we estimate the cooling rate to be of the order of $\SI{0.1}{\nano\kelvin/\milli \second}$.

\par

\section{Model limitations}

To reach our quantitative predictions we have made the following, assumptions:

\par

(i)~The WM interactions with the baths are assumed to be Markovian, i.e., devoid of memory effects such as back-propagation of excitations to their origin. The Markovian assumption is unquestionable in the non-degenerate regime. In the deeply degenerate regime, where only a few modes may be excited, non-Markovian recurrences may be encountered~\cite{rauer2018recurrences}. Such non-Markovianity is beyond the scope of the paper, but its effects may give rise to a most fascinating experimental regime. It must, however, be noted that the energy changes in the Otto cycle only depend on the four end points in Fig.~\ref{fig_cycle}b and are thus independent of the dynamics that connects these points. We note that this dependence of energy changes only on the end points holds since energy is a system variable. Importantly, in this cycle the heat transfer equals the energy change in the respective paths (since they are isochoric) and hence is path-independent. Generally, this is not the case, e.g., in the Diesel and Carnot cycles.

\par

(ii)~The discrete excitation spectrum of the bath has been neglected here, i.e., the local density approximation has been adopted. This is an excellent approximation for weak traps and high enough temperatures~\cite{pitaevskii1998bose,zambelli2000dynamic} but it may break down at extremely low temperatures, which is the range of possible non-Markovian dynamics.

\par

(iii)~We have neglected interactions between the bath modes, which is an excellent approximation even in the condensed regime, where the excitations are well-described by non-interacting Bogoliubov modes~\cite{dalfovo1999theory, rauer2018recurrences}. We note, however, that these interactions must be included in a fully quantitative study of the thermalization dynamics, which is beyond the scope of this paper.

\par

(iv)~The temperature of each bath has been assumed to be constant during each stroke [\cf Eq.~\eqref{eq_temperature_change}] as we have a temperature change that is less than $1\%$ of the temperature of the bath itself. We can verify in Fig.~\ref{fig_temperatures_above_Tcrit_h_Tc_700} that this approximation is well justified.

\par

(v)~For a simplified treatment of the WM, we have assumed that only two energy levels are accessible. Such a situation can be naturally realized in an optical trap, which is anisotropic and singles out the direction of propagation of the laser~\cite{grimm2000optical}.

\par

(vi)~Our treatment has assumed that thermal occupancy of higher modes can be neglected, which means that the energy-level spacing is always larger than the bath temperature. This assumption is valid for the chosen parameters in our numerical treatment. Although optimization of heat extraction from the cold bath may call for a WM which occupies a large number of modes, the entire setup would then be well described by a classical (rather than quantized) Otto cycle, similar to the description of Ref.~\cite{rossnagel2016single}.

\par

(vii)~The Otto cycle presumes an adiabatic change of the energy-level spacing in the WM so as to avoid quantum friction~\cite{kosloff2017quantum}. For strokes that are longer than $\SI{1}{\milli \second}$ this is indeed the case. Strokes involving a faster change of the trap frequency may be exploited in the future by employing shortcuts to adiabaticity~\cite{delcampo2014more}.

\par

(viii)~We have assumed weak coupling between the WM and the bath, consistently with the Markovian approximation. This assumption is well established within the context of the Fr\"ohlich polaron, which has been extensively investigated~\cite{grusdt2016new,hu2016bose,jorgensen2016observation,rentrop2016observation}. Notably, we have verified that the dimensionless coupling parameter, which describes the deformation of the bath density due to the coupling with the WM, is always much smaller than unity. The weak-coupling regime has also allowed us to neglect the energy cost of decoupling the WM from the bath and the corresponding entropy changes.

\par

(ix)~Finally, thermalization between the baths and the WM during the isochores has been ensured by considering time scales exceeding the relaxation time~\cite{myatt1997production}, as discussed in more detail below. In the future we may explore how such a thermal machine functions if thermalization is hampered (due to the occurrence of non-thermal fixed points~\cite{pruefer2018observation, eigen2018universal, erne2018universal}, the existence of dark states~\cite{gelbwaser2015power}, invariant subspaces~\cite{niedenzu2018cooperative} or many-body-localization~\cite{nandkishore2015many}).

\par

\section{Experimental challenges}

Having established the feasibility of the quantum gas refrigerator and its conceivable limitations we now turn our attention to possible experimental challenges. One challenge is the attainment of sufficiently fast thermalization of the WM with the baths, relying on strong interspecies interactions in Rb-Cs~\cite{spethmann2012dynamics} or Na-K~\cite{hartmann2019feshbach} mixtures. We can then estimate thermalization rates based on the results of Ref.~\cite{scelle2013motional} that apply to the proposed setup. With the assumed parameters, we obtain typical relaxation times on the order of a $\SI{1}{\milli\second}$, which are similar to the WM adiabaticity time scale.

\par

A fast enough relaxation is crucial in view of the finite lifetime of the atomic clouds, which is typically of a few seconds, and residual heat effects. The species-selective optical control of the WM has to be designed carefully; such that it provides a strong confinement of the WM but does not perturb the heat baths. For the proposed mixture of Rb-Cs the optical potential might be tuned to $\SI{880}{\nano\meter}$, which is the tune-out wavelength of Cs~\cite{leblanc2007species}, such that it does not create a dipole potential for the baths. Still, a confining potential with a waist of a few microns will cause an absorption rate of a few $\si{\hertz}$ within the WM. The produced heat per particle and per stroke is much smaller than the heat exchanged between the baths (Appendix~\ref{App:SponHeating}) and therefore negligible.

\par

The high density of atoms within the WM is unrealistic because of inelastic processes that would degrade the WM. Hence, it would be important to decompose the WM into a number of small and dilute WMs trapped in an optical lattice potential as sketched in Fig.~\ref{fig_experiment}. Cooperative effects may play a major role in this regime~\cite{scully2009collective,mazets2007multiatom,manatuly2019collectively}, a problem which has to be studied experimentally.

\par

The physical decoupling of the WM from the bath is implicitly assumed to be adiabatic. In the condensed regime the Bose-Einstein condensate is a superfluid and motion below the critical velocity does not create any excitations within the bath~\cite{jendrzejewski2014resistive}.

\par

For the non-degenerate bath, on the other hand, we derive in Appendix~\ref{App:BathMotion} the adiabatic decoupling condition according to which the WM has to move well below the velocity $u_\mathrm{a} = \omegaT\lambda_\mathrm{dB}$. For our assumed parameters, this would imply typical cycles times in the order of few tens $\SI{}{\ms}$, such that only few strokes would be possible. If the adiabatic decoupling condition is violated, the interaction between the  WM and bath is quenched in each cycle. The deposited energy is then estimated to be $\Delta E = g N_\mathrm{WM}n_\mathrm{bath}$, which is typically larger than the extracted heat. Thus the adiabatic decoupling condition imposes severe restrictions on the Otto cycle operation in the non-degenerate bath regime. These restrictions may however be overcome through shortcuts to adiabaticity~\cite{torrontegui2013shortcuts,delcampo2014more,delcampo2019friction}. A fundamental solution to this problem is a continuous cycle scheme, wherein coupling and decoupling are totally absent, as we discuss in the next section.

\par

\section{Continuous cycles schemes}

\begin{figure}
  \centering
  \includegraphics[width=\columnwidth]{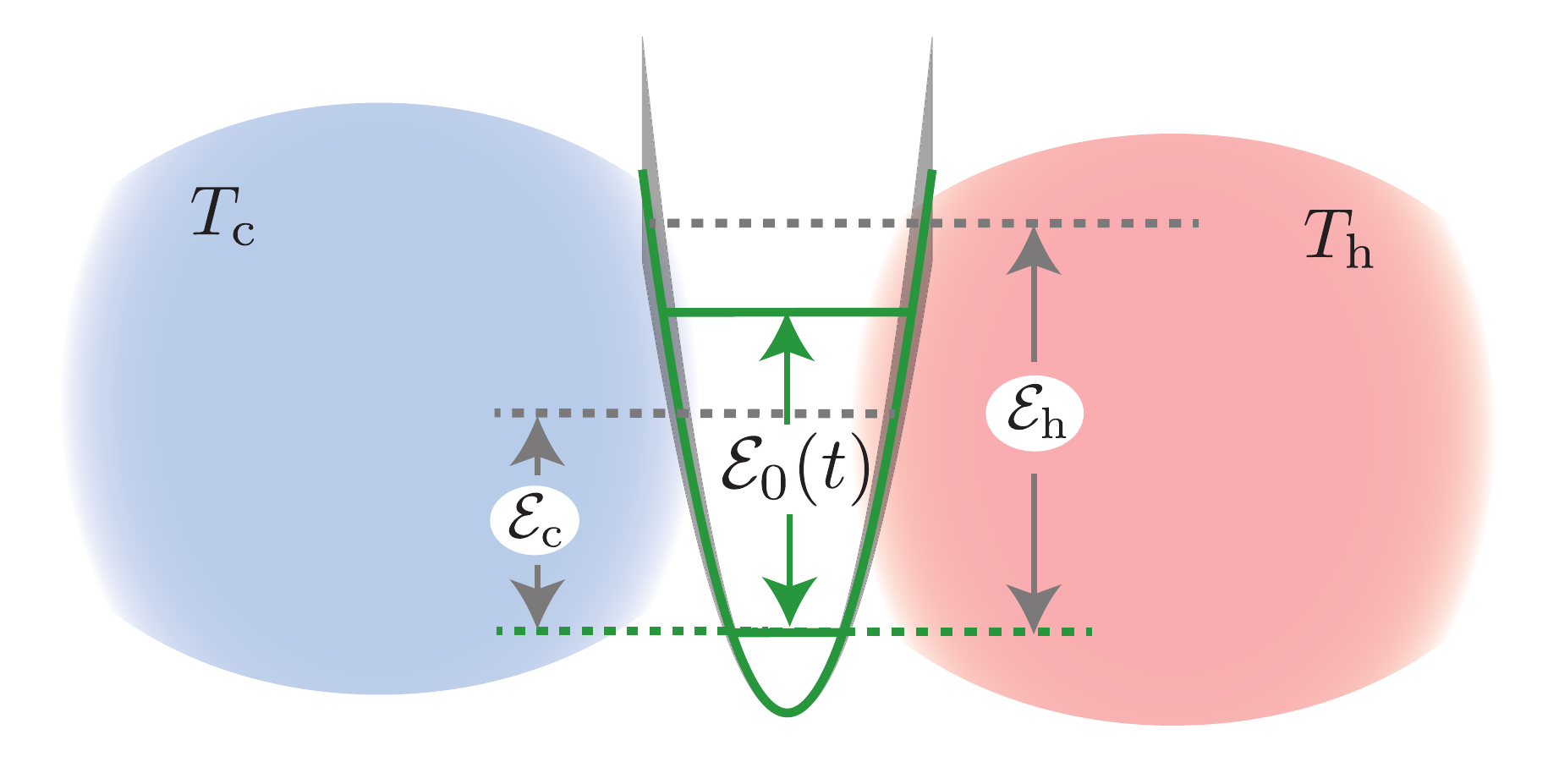}
  \caption{\textbf{Continuous cycle schemes}: The working medium is positioned in between the two baths. The periodic modulation of the energy spacing $\mathcal{E}_0(t)$ induces energy sidebands $\omegah$ and $\omegac$. The spectra of the two baths have to be engineered such that the working medium is stronger coupled at $\omegah$ ($\omegac$) to the hot (cold) bath.}\label{fig_conti}
\end{figure}

As an alternative to the Otto cycle, one may consider a different class of thermodynamic cycles, namely, continuous cycles wherein both the heat baths as well as the driving laser (the work reservoir) are simultaneously coupled to the WM~\cite{kolar2012quantum,gelbwaser2013minimal,kosloff2013quantum,alicki2014quantum,gelbwaser2015thermodynamics,brandner2016periodic,mukherjee2016speed}. A possible setup is visualized in Fig.~\ref{fig_conti}.

Instead of coupling the WM to the respective baths in two different (temporarily-separated) isochoric strokes at respective WM energy-level spacings $\omegac$ and $\omegah$, a spectral separation of the two baths has been proposed, such that blue (red) sidebands of the mean trap frequency $\omega_\mathrm{T}$ mainly couple to the hot (cold) bath~\cite{gelbwaser2013minimal,alicki2014quantum}, similarly to the Stokes and anti-Stokes sidebands in laser cooling~\cite{wineland1979laser}.

\par

The salient advantage of such a continuous cycle is that neither (de)coupling of the WM to or from the baths nor its translation between the baths are necessary. Such a cycle is not restricted by the bottleneck of adiabatic strokes and thus may be much faster than the Otto cycle and thereby potentially increase the cooling power. On the other hand, the required spectral separation of the two baths is non-trivial, yet it may be achieved through internal-state manipulations of the baths confined in optical lattices.

\par

The theoretical description of continuous cycles, promising as they may be, requires a more detailed knowledge of the concrete setup than the Otto cycle. In particular, the Floquet analysis~\cite{szczygielski2014application} of the Markovian master equation governing the evolution of the WM in a thermal gas (derived in Ref.~\cite{lewenstein1995master} for a constant trap frequency) for a periodically-driven trap frequency may require fine-tuning of the frequencies involved. Further, modulating the laser intensity only varies the ``spring constant'' of the WM trap but not its mass, which results in a non-diagonal modulation~\cite{kosloff2017quantum} with non-trivial time-dependencies of the excitation and de-excitation operators, which are not present in the previously-studied two-level atom case~\cite{gelbwaser2013minimal}. Finally, the validity of the Markovian description (which is at the heart of Refs.~\cite{geva1995relaxation,gelbwaser2013minimal,kosloff2013quantum,alicki2014quantum,gelbwaser2015thermodynamics,brandner2016periodic}) must be carefully re-analysed for such a cycle\footnote{First results on a Markovian master equation below threshold for a constant trap frequency can be found in the PhD thesis of Raphael Scelle~\cite{scellephd}.}. Concrete predictions require a detailed knowledge of the bath spectra at the respective sideband frequencies. An experimental implementation of the continuous cycle would hence provide crucial inputs to its theoretical descriptions.

\par

\section{Summary}

We have adopted the harmonic Otto cycle of Ref.~\cite{kosloff2017quantum} to propose a novel quantized refrigeration scheme for a cold bosonic-atom species, considered as a trapped cold bath. The scheme is based on the coupling of this cold bath to another trapped atomic species that acts as a working medium (WM), which transfers heat from the cold bath to a hotter bath. The process is enabled by the work invested in changing the energy-level spacing of the trapped WM by modulating the intensity of the trapping field. Our analysis has shown the feasibility of Otto-cycle refrigeration (especially when assisted by shortcuts to adiabaticity) of a typical bosonic-atom (e.g., Cs) cloud from a thermal regime to a deeply degenerate Bose--Einstein condensate while conserving the number of atoms in the cloud. This task, shown here to be at hand, is unfeasible by other means.

\par

The proposed scheme promises to be a fundamentally new avenue for cooling as it does not rely on radiative thermalization and is not recoil-limited. Continuous versions of the proposed cycles are potentially even more promising, because they are expected to be free of adverse non-adiabatic effects~\cite{kolar2012quantum,gelbwaser2013minimal,kosloff2013quantum,gelbwaser2015thermodynamics}. On the other hand, the continuous version requires parameter choice and optimization that may be easier to achieve experimentally than theoretically.

\par

The minimal temperature achievable by the proposed refrigeration is an open issue. The present minimum, Eq.~\eqref{eq_otto_Tc_min}, is only restricted by our technical ability to ramp the confining laser intensity up and down, so that further progress is expected in this respect. The more fundamental issue is the ability to cool down to such low temperatures that the Markov and possibly the Born approximations break down. The cooling speed may be affected at such temperatures in ways still unknown. Related studies~\cite{erez2008thermodynamic,gordon2009cooling,alvarez2010zeno,whitney2018nonmarkovian} indicate a plethora of fascinating effects in this non-Markovian domain.

\par

The open fundamental issues outlined above suggest that the proposed refrigerator may be a platform for studying principal aspects of quantum thermodynamics in hitherto inaccessible regimes.

\par

\section*{Acknowledgements}

We are grateful for fruitful discussions with David Gelbwaser-Klimovsky. W.\,N.\ acknowledges support from an ESQ fellowship of the Austrian Academy of Sciences (\"OAW). G.\,K. acknowledges support by the ISF. G.\,K, I.\,M. and F.\,J. jointly acknowledge the DFG support through the project FOR 2724. I.\,M. acknowledges support by the Wissenschafts- und TechnologieFonds (WWTF) project No. MA16-066 (``SEQUEX''). F.\,J.\ acknowledges support by the DFG (Project-ID 377616843), the Excellence Initiative of the German federal government and the state governments—funding line Institutional Strategy (Zukunftskonzept): DFG project number ZUK49/\"U.

\appendix

\section{Heat and work in the quantized Otto cycle}\label{app_otto}

The respective changes in the energy of the WM during the four strokes of the Otto cycle from Fig.~\ref{fig_cycle} are~\cite{abah2016optimal}
\begin{subequations}\label{eq_otto_energies}
  \begin{alignat}{3}
    &\Delta E_1&&=\ew{H_\indexc}_B-\ew{H_\indexc}_A&&\eqcolon Q_\indexc\\
    &\Delta E_2&&=\ew{H_\indexh}_C-\ew{H_\indexc}_B&&\eqcolon W_\mathrm{in}\\
    &\Delta E_3&&=\ew{H_\indexh}_D-\ew{H_\indexh}_C&&\eqcolon Q_\indexh\\
    &\Delta E_4&&=\ew{H_\indexc}_A-\ew{H_\indexh}_D&&\eqcolon W_\mathrm{out}.
  \end{alignat}
\end{subequations}
Here $H_\mathrm{c,h}=\mathcal{E}_\mathrm{c,h}\splus\sminus$ are the respective Hamiltonians of the WM during the cold and hot strokes with $\omegac \leq \omegah$. The state of the WM does not change during the adiabatic strokes ($B\rightarrow C$ and $D\rightarrow A$), i.e., the ratio $\mathcal{E}/T$ remains constant. Therefore, the thermal states at the four end points of the cycle are
\begin{subequations}
  \begin{alignat}{4}
    &\rho_A&&\equiv\rho_D&&=\exp[- H_\indexh/(\kB\Th)]/Z_\indexh&&\eqcolon\rho_\indexh\\
    &\rho_B&&\equiv\rho_C&&=\exp[- H_\indexc/(\kB\Tc)]/Z_\indexc&&\eqcolon\rho_\indexc.
  \end{alignat}
\end{subequations}

\par

The work ($W$) and heat ($Q$) contributions~\eqref{eq_otto_energies} then evaluate to
\begin{subequations}\label{eq_otto_tls}
  \begin{align}
    Q_\indexc&=\omegac(\nbar_\indexc-\nbar_\indexh)\\
    W_\mathrm{in}&=(\omegah-\omegac)\nbar_\indexc\\
    Q_\indexh&=\omegah(\nbar_\indexh-\nbar_\indexc)\\
    W_\mathrm{out}&=(\omegac-\omegah)\nbar_\indexh,
  \end{align}
\end{subequations}
where ($i\in\{\indexc,\indexh\}$)
\begin{equation}
  \nbar_i\coloneq \ew{\splus\sminus}_{\rho_i}=\left[\exp\left[\mathcal{E}_i/(\kB T_i)\right]+1\right]^{-1}
\end{equation}
is the thermal occupation of the excited level of the two-level WM.

\section{Heat capacity of a Bose gas above and below threshold}\label{App:BecProperties}

The heat capacity of a BEC at constant volume in a three-dimensional harmonic trap reads~\cite{pethickbook}
\begin{equation}\label{eq_CV_BEC}
  C_\mathrm{V}=
  \begin{cases}
    3N_\mathrm{at}\kB\left[1+\frac{\zeta(3)}{8}\left(\frac{T_\mathrm{crit}}{T}\right)^3\right] & T>T_\mathrm{crit} \\
  12\frac{\zeta(4)}{\zeta(3)}N_\mathrm{at}\kB\left(\frac{T}{T_\mathrm{crit}}\right)^3 & T<T_\mathrm{crit}
  \end{cases}
  ,
\end{equation}
where the critical temperature for condensation is~\cite{pethickbook}
\begin{equation}\label{eq_Tcrit}
  T_\mathrm{crit}=\frac{\hbar\omega_\mathrm{T}N_\mathrm{at}^{1/3}}{\kB [\zeta(3)]^{1/3}}.
\end{equation}
Here $N_\mathrm{at}$ denotes the atom number in the bath, $\omega_\mathrm{T}$ the trap frequency and $\zeta$ is the Riemann $\zeta$-function.

\section{Heating due to spontaneous emission}\label{App:SponHeating}

The optical confinement of the WM induces spontaneous emission events during which the WM atoms gain a recoil energy $E_\mathrm{R}=h^2/(2\mWM \lambdaWM^2)$, where $h$ is the Planck constant, $\mWM$ the mass of the atoms and $\lambdaWM$ the wavelength of the emitted photons ($\SI{780}{\nano\meter}$ in the case of Rb and hence $E_\mathrm{R}/\kB \approx \SI{180}{\nano\kelvin}$). The produced heat per atom in the WM can be estimated as $Q_\mathrm{sp} = E_\mathrm{R} \gammaWM \tauStroke$, where $\gammaWM$ is the rate of spontaneous emission (typically a few $\si{Hz}$) and $\tauStroke$ the duration of each stroke (on the order of $\si{\milli \second}$). Therefore, we estimate $Q_\mathrm{sp}\kB\sim\textrm{a few } \si{\nano\kelvin}$ per stroke, which typically is much smaller than the exchanged heat in Eq.~\eqref{eq_otto_tls}.

\begin{widetext}

\section{Motion of the WM in the non-degenerate bath}\label{App:BathMotion}

Here we evaluate the effect of the non-adiabatic movement of a compact WM through a bath. We assume that the bath to be composed of a non-degenerate and (almost) ideal gas and strive to estimate the transferred energy $\langle \Delta E \rangle $ per single atom. For the sake of simplicity, we consider a 1D harmonic trap with the frequency $\omegaT $. The typical inverse length scale for an atom of mass $m$ is $\alpha =\sqrt{m\omegaT /\hbar }$. The energy spectrum is $E_n=\hbar \omegaT n$, $n=0,1,2,\dots $

\par

Since we are interested in the case of small transferred energy, we apply the lowest-order perturbative expression for the transition amplitude of an atom being driven from the state $|n_0\rangle $ to the state $|n\rangle $ under the time-dependent perturbation $V(z,t)$,
\begin{equation}
  {\cal A}_{n_0\rightarrow n} =-\frac i\hbar \int _{-\infty }^\infty dt\, e^{i(E_n-E_{n_0})t/\hbar }\langle n|V|n_0\rangle .
  \label{A1}
\end{equation}
Then the transferred energy per atom is
\begin{equation}
  \langle \Delta E \rangle ={\cal Z}^{-1}\sum _{n_0=0}^\infty \sum _{n=0}^\infty (E_n-E_{n_0})e^{-\beta E_{n_0}}
  |{\cal A}_{n_0\rightarrow n}|^2 ,
  \label{E1}
\end{equation}
where $\beta $ is the inverse temperature and ${\cal Z}=1/( 1-e^{-\beta \hbar \omegaT })$ the partition function.
Slightly reorganizing the summation, we see that
\begin{equation}
  \langle \Delta E \rangle ={\cal Z}^{-1}\sum _{n_0=0}^\infty \sum _{\, n=n_0+1}^\infty (E_n-E_{n_0})(e^{-\beta E_{n_0}}-e^{-\beta E_n})
  |{\cal A}_{n_0\rightarrow n}|^2
  \label{E2}
\end{equation}
is always positive for a bath in thermal equilibrium.

\par

For the purpose of evaluation of the non-adiabatic effects it is convenient to approximate the potential of interaction of a bath atom with the working medium. The contact interaction between the bath atoms and the working medium reads $E=g_\mathrm{IB}\int \dd z~n_\mathrm{WM}(z-ut) n_\mathrm{th}(z)$.
For a very small working medium it looks like a potential and the bath density is  in its region, such that
\begin{equation}
  V(z,t)=V_0 \delta (z-ut) \mbox{ with }  V_0= g_\mathrm{IB}\NWM
  \label{V1}
\end{equation}
Here we assume that the working medium passes through the bath at the constant velocity $u$.

\par

Now we readily calculate
\begin{equation}
  {\cal A}_{n_0\rightarrow n}=-\frac {iV_0}{u\hbar}\int _{-\infty }^\infty dz\, \psi _n^*(z)e^{i\omegaT (n-{n_0})z/u}\psi _{n_0}(z),
  \label{A2}
\end{equation}
where $\psi _{n,\, n_0}$ are eigenfunctions of the harmonic oscillator. Expressing the coordinate $z$ via the creation and annihilation
operators $\hat a^\dag $ and $\hat a$, respectively, we recognize that ${\cal A}_{n_0\rightarrow n}$ is proportional
to the matrix element of the displacement operator $\exp [i\lambda _n(\hat a^\dag +\hat a)]$ with
$\lambda _n=\omegaT (n-n_0)/(\sqrt{2}u\alpha )$ in the Fock basis. These matrix elements are known, see, e.g., Ref.~\cite{wunsche1991displaced}.
Since, according to Eq.~\eqref{E2}, $n>n_0$, we get
\begin{equation}
  \langle n|e^{i\lambda _n(\hat a^\dag +\hat a)} |n\rangle = i^{n-n_0}e^{-\lambda _n^2/2} \lambda _n^{n-n_0}
  \sqrt{ \frac {n_0!}{n!}} L _{n_0}^{n-n_0}(\lambda _n^{2}),
  \label{ME1}
\end{equation}
where $L _{n_0}^{n-n_0}(x)$ is an associated Laguerre polynomial. This reduces Eq.~\eqref{E2} to
\begin{equation}
  \langle \Delta E \rangle =(1-e^{-\beta \hbar \omegaT })\hbar \omegaT \left( \frac {V_0}{\hbar u}\right) ^2 \sum_{n_0=0}^\infty
  \sum _{l=1}^\infty e^{-\beta \hbar \omegaT n_0}l (1-e^{-\beta \hbar \omegaT l})
  \zeta ^{2l}l^{2l}e^{-\zeta ^2l^2}\frac {n_0!}{(n_0+l)!} [L_{n_0}^l(\zeta ^2l^2)]^2,
  \label{E3}
\end{equation}
where $\zeta =\omegaT /(\sqrt 2 \alpha u)$. The sum over $n_0$ in Eq.~\eqref{E3} can be taken using the Hille--Hardy formula
\cite{batemanbook}, thus yielding
\begin{equation}
  \langle \Delta E \rangle =2\hbar \omegaT \left( \frac {V_0}{\hbar u}\right) ^2
  \sum _{l=1}^\infty l \sinh (\beta \hbar \omegaT /2) \exp [ - \zeta ^2l^2\coth (\beta \hbar \omegaT /2)]
  I_l[\zeta ^2l^2 /\sinh (\beta \hbar \omegaT /2)]  ,
  \label{E4}
\end{equation}
where $I_l(x)$ is the modified Bessel function.

\par

We work in the high-temperature limit,
\begin{equation}
  \beta \hbar \omegaT \ll 1 .
  \label{HT1}
\end{equation}
In this limit the adiabaticity condition requires that the velocity of the working medium in the bath does not exceed a certain value,
\begin{align}
  u\ll u_\mathrm{a} &\equiv \frac{\omegaT}{\alpha} \sqrt{\frac {\beta \hbar \omegaT }{8}}\\
  u_\mathrm{a}&= \omegaT \lambda_\mathrm{dB}\mbox{ with }\lambda_\mathrm{dB}= \sqrt{\frac {\beta \hbar^2 }{8m}},
                \label{UA1}
\end{align}
where $\omegaT$ sets the typical gap and and $\lambda_\mathrm{dB}$ the typical distance. Note that condition~\eqref{UA1} is much more restrictive that the smallness of the oscillation period compared to the passage time.

\par

To derive this condition, we assume first a less stringent inequality, $u\ll \omegaT /( \alpha \sqrt{\beta \hbar \omegaT })$, which means that the time of the passage of the working medium across the bath of the thermal radius $\sim 1/( \alpha \sqrt{\beta \hbar \omegaT })$ is much longer than the trap oscillation period. Then we can use the asymptotic form $I_l(x) \approx e^x/\sqrt{2\pi x}$ for $x\rightarrow +\infty $. Equation~\eqref{E4} is then reduced to
\begin{equation}
  \langle \Delta E \rangle =2\hbar \omegaT \left( \frac {V_0}{\hbar u}\right) ^2
  \sum _{l=1}^\infty l \frac{\beta \hbar \omegaT}{2} \exp \left( - \frac{\zeta ^2l^2\beta \hbar \omegaT}{4}\right) \frac{\sqrt{\beta\hbar\omegaT}}{2\sqrt{\pi}\zeta l}  ,
  \label{E5}
\end{equation}
where condition~\eqref{HT1} has been used. In the adiabatic limit given by Eq.~\eqref{UA1} we obtain
\begin{align}
  \langle \Delta E \rangle &=\frac 2{\sqrt \pi } \beta(\alpha V_0)^2 \frac {u_\mathrm{a}}u \exp \left( -\frac {u_\mathrm{a}^2}{u^2}\right).
                             \label{R1}
\end{align}
Under the latter condition the decreasing exponent and the increasing modified Bessel function almost compensate each other, hence the exponential suppression of the heating rate requires further slowing down of the motion of the working medium inside the bath. The strong exponential dependence (not a power-law one) of the r.h.s.\ of Eq.~\eqref{R1} on the adiabaticity parameter indicates that an adiabaticity condition similar to Eq.~\eqref{UA1} should hold also in 3D.

\end{widetext}

\end{document}